\documentclass[a4paper,11pt]{article}
\pdfoutput=1
\usepackage{amsmath}
\usepackage{amssymb}
\usepackage{color}
\usepackage{cite}
\usepackage{hyperref}
\usepackage{graphicx}
 \usepackage{leftidx}
 \usepackage{subfig}

\numberwithin{equation}{section}   

\def \be {\begin{equation}}
\def \ee {\end{equation}}
\def \ba {\begin{array}}
\def \ea {\end{array}}
\def \bea{\begin{eqnarray}}
\def \eea{\end{eqnarray}}

\def \d {\delta}
\def \D {\Delta}

\begin{document}

\title{\textbf{An Interpretation for the Equivalence of Two Holographic Computations of the Butterfly Velocity with the Canonical Formalism of Gravity}}
\author{Feiyu Deng$^{1,2}$\footnote{dengfy@itp.ac.cn}\,
Xiao-Shuai Wang$^{1,2}$\footnote{wangxiaoshuai@itp.ac.cn}\,
Jie-qiang Wu$^{1,2}$\footnote{jieqiangwu@itp.ac.cn}
}
\date{}

\maketitle

\begin{center}
{\it
$^{1}$Institute of Theoretical Physics, Chinese Academy of Sciences, Beijing 100190, China \\
$^{2}$School of Physical Sciences, University of Chinese Academy of Sciences, \\ Beijing 100049, China 
}
\vspace{10mm}
\end{center}

\begin{abstract}
In this paper, we revisit the equivalence of two holographic computations of the butterfly
velocity: the computation with the shock wave solution and the computation with the
entanglement wedge reconstruction. We provide an interpretation for the equivalence of
the two computations with the canonical formalism of gravity.

Specifically, by taking use of the canonical formalism, we reformulate both computations
into the ones with a similar form. Here, in both reformulated computations, the butterfly
velocity is computed from applying a given set of initial data into the constraint equations.
And the sets of initial data of both computations have a similar structure.

We then interpret the equivalence of the two computations as from the similar form of
the reformulated computations.
\end{abstract}

\baselineskip 18pt
\thispagestyle{empty}
\newpage
\tableofcontents
\section{Introduction}\label{sec:intro}

Chaos has increasingly been recognized as a fundamental probe in the exploration of quantum gravity, offering insights that connect black hole physics, holography, and quantum information theory \cite{Susskind:2013aaa,Shenker:2013pqa,Hayden:2007cs,Narovlansky:2025tpb}. A particularly striking manifestation of this probe is the butterfly effect, which characterizes the exponential growth of small perturbations. The quantitative description of the butterfly effect relies on the notions of the Lyapunov exponent and the butterfly velocity \cite{Shenker:2014cwa,Roberts:2016wdl}, which describe the temporal and spatial evolution of the perturbation front, often referred to as the butterfly cone \cite{Mezei:2019dfv}.

At the classical level, the butterfly effect and the characterization of chaos through Lyapunov exponents have been extensively studied, as systematically collected in some textbooks such as \cite{Lichtenberg:1992}. The essential mechanism for classical chaos is the sensitivity of trajectories in phase space to tiny variations in initial data. 

At the quantum level, there are no classical trajectories in Hilbert space, which leads to the absence of a unified definition of chaos \cite{Berry1977}. Nevertheless, the out-of-time-order correlator (OTOC) has emerged as a widely used diagnostic of quantum chaos. OTOCs circumvent a naive approach that considers sensitivity to quantum states\footnote{If one naively considers the overlap between an initial state and some perturbed state, it is always constant due to the unitarity of time evolution \cite{Cotler:2017myn}.}, instead being presented as operator growth in the Heisenberg picture \cite{Larkin1969}.

In addition to OTOCs, various other quantitative methods have been developed to investigate quantum chaos in the context of holography. Among various approaches, three have become particularly prominent in holographic studies: the shockwave method \cite{Shenker:2013pqa,Shenker:2013yza,Roberts:2014isa}, the entanglement wedge reconstruction method \cite{Mezei:2016wfz,Mezei:2016zxg}, and the pole-skipping mode analysis \cite{Grozdanov:2017ajz,Blake:2017ris,Blake:2018leo}. For a given perturbation at the asymptotic boundary, these approaches provide distinct holographic realizations of its propagation. In the shockwave method, it corresponds to a particle (or wavepacket) falling into the black hole, which backreacts significantly on the spacetime due to the large blueshift. In the entanglement wedge reconstruction method, it corresponds to a particle (or wavepacket) falling into the black hole, which drives the growth of the entanglement wedge associated with the minimal one of RT surfaces enclosing it. Unlike the shockwave and entanglement wedge reconstruction methods, pole-skipping mode does not explicitly correspond to a localized bulk excitation, but is instead revealed through the retarded correlator.

Despite their seemingly distinct formulations, growing evidence \cite{Mezei:2016wfz,Ahn:2019rnq,Blake:2021hjj,Grozdanov:2018kkt,Dong:2022ucb,Wang:2022mcq,Ning:2023ggs,Baishya:2024gih,Chua:2025vig,Ahn:2025exp,Chakraborty:2025crb} suggests that these approaches are in fact equivalent in computing the butterfly velocity. This equivalence was first pointed out in \cite{Mezei:2016wfz}, where the butterfly velocity obtained from shockwave and entanglement wedge reconstruction was shown to coincide in Einstein gravity, up to four-derivative corrections. Subsequently, this equivalence was extended to $f$(Riemann) theories in \cite{Dong:2022ucb}. More recently, the equivalence was further extended to include the pole-skipping phenomenon, as discussed in \cite{Chua:2025vig}. In parallel, the full or partial equivalence among these three approaches has been tested in a variety of specific backgrounds, such as hyperbolic black holes \cite{Ahn:2019rnq}, rotating black holes \cite{Blake:2021hjj}, asymptotically Lifshitz black hole \cite{Baishya:2024gih}.

In this paper, we aim to supplement the existing understanding of the equivalence between the shockwave and entanglement wedge reconstruction methods in computing butterfly velocity with Hamiltonian formalism. Owing to the pivotal role of the Hamiltonian formalism in elucidating gravitational dynamics and canonical quantization \cite{Dirac:1958sc,Arnowitt:1962hi}, it is natural to expect that it will uncover further structures underlying this equivalence. More importantly, this perspective resonates with the classical picture of chaos, where chaos is probed through the sensitivity to initial conditions.

To employ the Hamiltonian formalism, we need to foliate the spacetime with the ADM formulation in \cite{Arnowitt:1962hi}, and identify appropriate diffeomorphism-invariant observables that generate the evolution of initial data, where the constraint equations hold. In contrast to directly solving the equations of motion, which is commonly adopted in previous studies, we instead solve the constraint equations to obtain the shockwave profile $h(x)$ that encode the butterfly velocity in the shockwave method. Moreover, we adopt a method \cite{Bousso:2020yxi} based on the constraint equations to determine the location of the extremal surface, thereby unveiling the RT profile $\rho(x)$.  This approach to locating the extremal surface differs from the commonly used method, i.e. extremizing the entanglement entropy functional, and a new method from replica manifolds \cite{Chua:2025vig}. It is worth noting that, in order to discuss the equivalence between the shockwave and entanglement wedge reconstruction methods, we need to take a double shockwave \cite{Shenker:2013yza,Penington:2025hrc} into consideration in the shockwave method.

In this paper, we consider a general $(d+1)$-dimensional eternal black hole in $AdS_{d+1}$ \cite{Maldacena:2001kr} as a concrete example. We first represent a single shockwave backreaction on the spacetime as a kind of evolution of initial data on the Cauchy surface, which yields constraints equations satisfied by shockwave profile $h(x)$. We take the HRT area \cite{Hubeny:2007xt} as the observable of interest, due to its significance in holography \cite{Ryu:2006bv,Ryu:2006ef,Jafferis:2014lza,Jafferis:2015del} and well-understood action on the initial data \cite{Bousso:2020yxi,Kaplan:2022orm,ongoing1}. The HRT area generates a Hamiltonian flow \cite{Kaplan:2022orm,ongoing1}, which gives rise to constraints equations satisfied by RT profile $\rho(x)$. To match two methods, we introduce a double shockwave setup to ensure that the induced metric on the Cauchy surface holds. Our result reveals that the shockwave profile $h(x)$ and the RT profile $\rho(x)$ obey identical equations, highlighting underlying equivalence of two butterfly velocity computations.

The plan for the rest of the paper is as follows. In section \ref{sec:butterfly v}, we briefly review two holographic calculations of the butterfly velocity. In section \ref{sec:Hamiltonian}, we derive the butterfly velocity with Hamiltonian formalism and demonstrate the equivalence between two methods. In section \ref{sec:discussion}, we discuss the results and comment on potential future directions.

\section{Butterfly velocity in holography}\label{sec:butterfly v}
In this section, we briefly review two holographic calculations of the butterfly velocity, the shockwave method and entanglement wedge reconstruction method. We refer the reader to \cite{Dong:2022ucb,Chua:2025vig} for a more detailed review of the butterfly velocity and its various holographic formulations. 

The butterfly velocity was originally introduced as a measure of the spatial rate at which a local perturbation (operator size \cite{Dong:2022ucb}) grows and quantum information scrambles, also viewed as an effective Lieb-Robinson bound under thermal states \cite{Roberts:2016wdl}. Before reviewing two holographic methods for computing the butterfly velocity, we first recall how it can be extracted from out-of-time-ordered correlators (OTOCs), particularly in strongly coupled quantum systems. 

We introduce OTOCs\footnote{Here in deriving the second line of (\ref{OTOC}), we have assumed the two operators to be unitary. As only the second term in (\ref{OTOC}) contributes nontrivially, we do not distinguish between (\ref{OTOC}) and the second term in (\ref{OTOC}) when referring to OTOCs.} by the commutator of two local operators $V$ and $W$ 
\begin{align}\label{OTOC}
C_\beta(t_0,x)=&\langle[V(0,x),W(-t_0,0)]^\dagger[V(0,x),W(-t_0,0)]\rangle_{\beta}\notag\\
=&2-2 \operatorname{Re}\langle V(0,x)^{\dagger} W\left(-t_0,0\right)^{\dagger} V(0,x) W\left(-t_0,0\right)\rangle_\beta,
\end{align}
where $\beta$ represents the inverse temperature of the thermal state. Physically, it can be viewed as inserting an operator $W$ at $(-t_0,0)$  and subsequently probing it with an operator $V$ at $(0,x)$.

In certain quantum systems, particularly strongly coupled ones, OTOCs exhibit exponential behavior in the scrambling region $|x|\leq v_B\left(t_0-t_*\right)$ around scrambling time $t_*$ \cite{Roberts:2016wdl}, i.e.
\be C_\beta\left(t_0,x\right) \sim e^{\lambda_L\left(t_0-t_*-|x| / v_B\right)}, \ee
where $\lambda_L$ denotes the Lyapunov exponent, which governs scrambling at a fixed spatial location. Here the scrambling time $t_*$ can be determined from the moment when $C_\beta$ at the spatial origin to be $\mathcal{O}(1)$.

\begin{figure}[h]
  \centering
  \includegraphics[width=7cm]{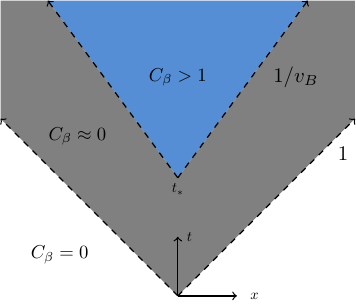}\\
  \caption{Interpretation of the butterfly cone. We denote the scrambling region by blue region, with the boundary $C_\beta\sim\mathcal{O}(1)$.}\label{Fig:butterfly cone}
\end{figure}

Furthermore, the time evolution of scrambling region can be interpreted as the butterfly cone \cite{Mezei:2019dfv}, shown in Fig. \ref{Fig:butterfly cone}. The boundary of butterfly cone is determined by $C_\beta\sim\mathcal{O}(1)$. It provides an intuitive depiction of how the scrambling region expands, with the butterfly velocity defining its rate of growth. In general, the butterfly velocity is bounded by the speed of light, meaning that it lies inside the light cone.

For simplicity, in the following we take $t_*=0$. We next briefly review two holographic computations of the butterfly velocity by considering a general $(d+1)$-dimensional eternal black hole
\be\label{eternal BH} ds^2=-a(uv)dudv + r(uv)^2\delta_{ij}dx^idx^j, \ee
where $u$ and $v$ denote null directions, $x^i$ denote transverse directions, in Kruskal coordinates. Here for notational convenience, we write the $ii$ component of the metric as $r(uv)^2$, which simplifies the expressions in the subsequent calculations.
An example of (\ref{eternal BH}) is a planar black hole \cite{Roberts:2014isa}
\be\label{planar BH} ds^2=-f(r)dt^2+\frac{dr^2}{f(r)}+r^2\delta_{ij}dx^idx^j, \ee
which corresponds to a finite temperature and infinitely large system in boundary field theory, where 
\be f(r) = r^2 - \frac{r_H^d}{r^{d-2}},\ee 
with 
\be \beta=\frac{4\pi}{d}\frac{1}{r_H}. \ee
We can check it satisfies the form of (\ref{eternal BH}) by following coordinate transformation
\begin{align}
u=&e^{\frac{2\pi}{\beta}(t + r_*)}\notag\\
v=&-e^{\frac{2\pi}{\beta}(-t + r_*)},
\end{align}
where 
\be dr_*=\frac{dr}{f(r)}. \ee

\subsection{The shockwave method}
\begin{figure}[h]
  \centering
  \includegraphics[width=6cm]{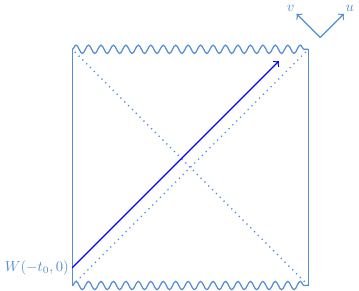}\\
  \caption{An infalling particle with large $t_0$.}\label{Fig:infalling pt}
\end{figure}

In this subsection, we review the shockwave method in computing butterfly velocity. In the shockwave method, inserting a local operator \(W(-t_0,0)\) on the boundary is equivalent to creating an infalling particle (energy packet) near the asymptotic boundary into the black hole at an early time as shown in Fig. \ref{Fig:infalling pt}.

We take $t_0$ to be large enough such that this localized shockwave contributes to the stress tensor with only $uu$ component nontrivial \cite{Dray:1984ha,Roberts:2014isa}, i.e.,
\be\label{dT} \d T_{\mu\nu}\sim e^{\frac{2\pi}{\beta}t_0}\d (u)\d (x)\d_{\mu}^u\d_{\nu}^u, \ee
where  the prefactor $e^{\frac{2\pi}{\beta}t_0}$ denotes a significant boost arising from blue shift. It gives rise to a significant backreaction on the spacetime, which only manifests in the $uu$ component of metric \cite{Dong:2022ucb}
\be\label{singlesw} ds^2=-a(uv)dudv + r(uv)^2\delta_{ij}dx^idx^j+h(x)\delta(u)du^2, \ee
where we refer to $h(x)$ as shockwave profile, which characterizes the deformation of horizon. The shockwave profile $h(x)$ is closely related to the butterfly velocity $v_B$, as in most cases it takes the following form
\be\label{h} h(x)\sim\frac{1}{|x|^{\#}}e^{\frac{2\pi}{\beta}t_0-\mu|x|}, \ee
with the butterfly velocity
\be v_B=\frac{2\pi}{\beta\mu}. \ee
Therefore, determining the butterfly velocity amounts to solving the shockwave profile $h(x)$.

Conventionally, the shockwave profile is obtained by solving the equations of motion
\be \d E_{\mu\nu}=\#\d T_{\mu\nu}. \ee
For example, we apply (\ref{planar BH}) on this equation, yielding the nontrivial equation, shockwave equation
\be\label{shock eq} \left(-\frac{1}{2r_H^2}\delta^{ij}\partial_i\partial_jh(x)+\frac{1}{4}d(d - 1)h(x)\right)\d (u)=\# e^{\frac{2\pi}{\beta}t_0}\d (x)\d (u). \ee
Its solution takes the form of (\ref{h}), with
\be \mu=r_H\sqrt{\frac{d(d-1)}{2}}. \ee

However, in general systems, demonstrating the equivalence of two methods by directly solving the equations of motion is extremely cumbersome, for instance in $f$(Riemann) theories \cite{Dong:2022ucb}. In the section \ref{sec:Hamiltonian}, we attempt to circumvent the explicit solution of the equations of motion, and instead derive a special form of the equation satisfied by the shockwave profile $h(x)$.

\subsection{The entanglement wedge reconstruction method}
In this subsection, we review the entanglement wedge reconstruction method in computing butterfly velocity. The entanglement wedge reconstruction \cite{Almheiri:2014lwa} provides a viewpoint of subregion/subregion duality \cite{Liu:2025krl}, where a given subregion in the boundary CFT is dual to its entanglement wedge. It motivates us to represent the scrambling region of the boundary CFT in terms of the entanglement wedge \cite{Mezei:2016wfz}, where we restrict our attention to the minimal entanglement wedge, to avoid possible ambiguities.

\begin{figure}[h]
  \centering
  \includegraphics[width=15cm,height=8cm]{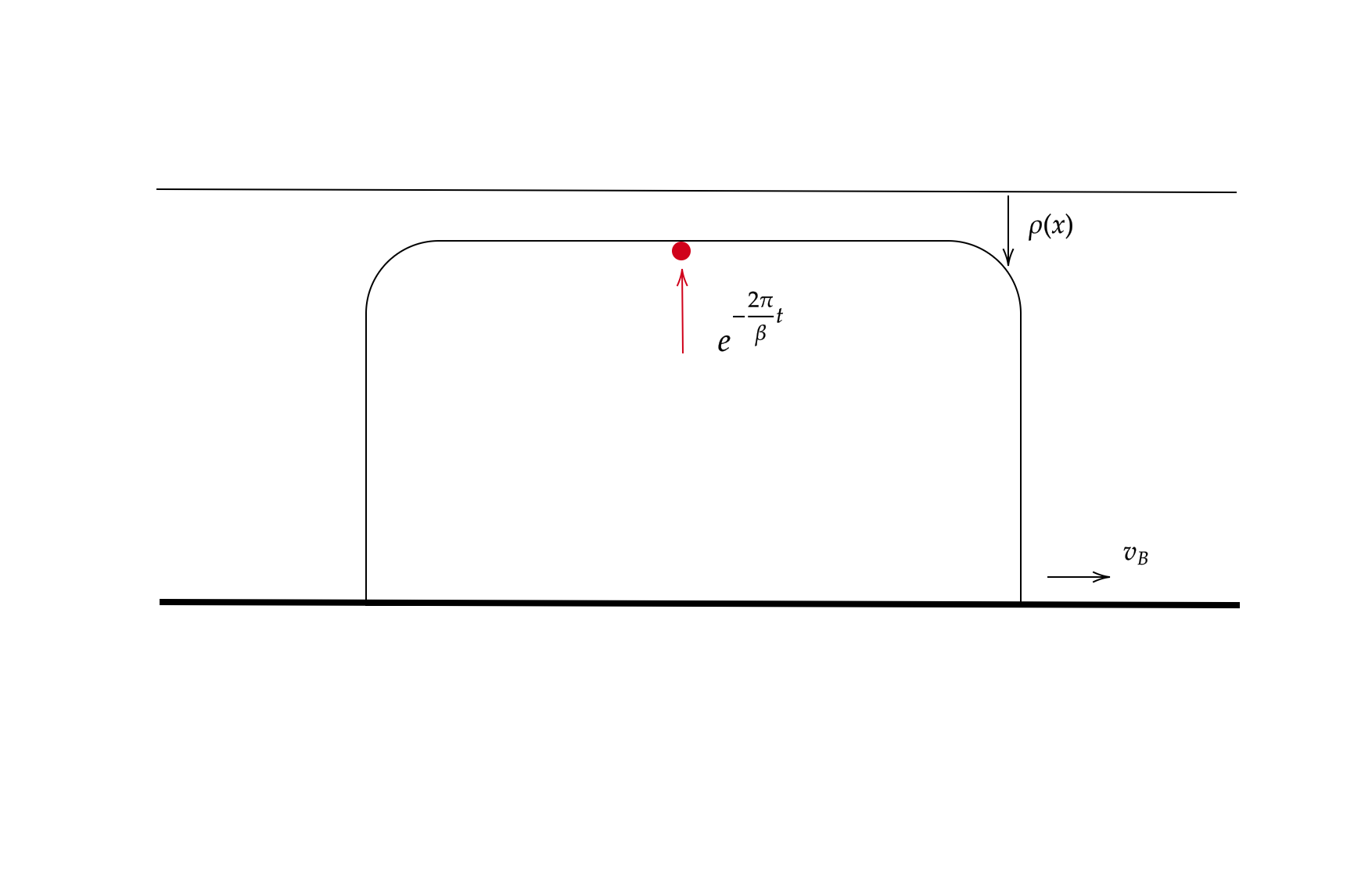}\\
  \caption{An illustration of the growth of entanglement wedge including an infalling particle.}\label{Fig:EW}
\end{figure}

We now elaborate on this construction in detail. Keeping that in mind the insertion of an operator in the boundary CFT corresponds to a particle falling into the black hole in the bulk, we can represent the growth of scrambling region of the boundary CFT as the growth of the minimal entanglement wedge, which is a HRT surface \cite{Hubeny:2007xt} including this infalling particle, as shown in Fig. \ref{Fig:EW}. 

At late times of the infalling process, when the infalling particle approaches the horizon, the shape of the HRT surface becomes less sensitive to the background geometry and takes the following form
\be\label{rho} \rho(t,x)\sim\frac{1}{|x|^{\#}}e^{-\frac{2\pi}{\beta}t+\widetilde{\mu}|x|}, \ee
in the near horizon region, with the center (the location of infalling particle)
\be \rho(t,x = 0)\sim e^{-\frac{2\pi}{\beta}t}. \ee
Here we refer to $\rho(t=\text{const},x)$ as the RT profile $\rho(x)$, in no essential difference from the RT profile appearing in \cite{Dong:2022ucb}. We then can extract the butterfly velocity from the growth of HRT surface near horizon region
\be \widetilde{v}_B=\frac{2\pi}{\beta\widetilde{\mu}}. \ee
Therefore, determining the butterfly velocity amounts to solving the RT profile $\rho(x)$ (the location of HRT surface).

An example corresponding to (\ref{planar BH}) is
\be -\frac{1}{2r_H^2}\delta^{ij}\partial_i\partial_j\rho(x)+\frac{1}{4}d(d - 1)\rho(x)=0, \ee
which differs from (\ref{shock eq}) by the absence of the source term. We also aim to employ the Hamiltonian formalism to patch up this mismatch.

The most common way to determine the location of the extremal surface is to extremize the entanglement entropy functional \cite{Dong:2022ucb}. In the following, we would like to adopt a less commonly used method to reformulate the location of the extremal surface, i.e., solving constraint equations \cite{Bousso:2020yxi}.

\section{Butterfly velocity from Hamiltonian formalism}\label{sec:Hamiltonian}
In this section, we derive butterfly velocity from Hamiltonian formalism. In particular, we seek to formulate the equations for shockwave profile $h(x)$ and RT profile $\rho(x)$. Additional details can be found in Appendix. \ref{sec:Appen}.

\subsection{Formulation of Hamiltonian formalism}
We first introduce the Hamiltonian formalism for AdS$_{d+1}$ gravity, following our previous setup in \cite{Wang:2023jfr}. We foliate the spacetime with a set of Cauchy surfaces $\Sigma_t$ with constant $t$. Given this foliation, we can reformulate the spacetime into the evolution of the set of initial data, i.e. the induced metric $\sigma_{ab}$ and the extrinsic curvature $K_{ab}$ on a Cauchy surface $\Sigma_{t_0}$. We can read out the induced metric $\sigma_{ab}$ from the foliation of metric
\be\label{foliation} g_{\mu\nu}=-N^2dt^2+\sigma_{ab}(dx^a+\beta^adt)(dx^b+\beta^bdt), \ee
where $\beta^a$ is shift vector, $N$ is the lapse function related to the covector $\tau_\mu=-N\d^t_\mu$ to the Cauchy surface. We can compute the extrinsic curvature by
\be\label{Kab} K_{ab}=\sigma_a^\mu\nabla_\mu\tau_b. \ee
A legal evolution of initial data should preserve the constraint equations $\mathcal{H}$ and $\mathcal{H}_a$\footnote{We will clarify this point in subsection \ref{subsec:EW Hamil}.}, i.e.,
\begin{align}
	\D\mathcal{H}\Big|_{(\D\sigma_{ab},\D K_{ab})}=&0\notag\\
	\D\mathcal{H}_a\Big|_{(\D\sigma_{ab},\D K_{ab})}=&0.
\end{align}

In the shockwave method, we reformulate the backreaction $\Delta g_{\mu\nu}$ into the evolution of initial data $\Delta\sigma_{ab}$ and $\Delta K_{ab}$. Considering the evolution of initial data need to preserve the constraint equations, we can get some equations for the shockwave profile $h(x)$.

In the entanglement wedge reconstruction method, we represent the initial data with the RT profile $\rho(x)$. In addition, we introduce a diffeomorphism-invariant observable, the HRT area \cite{Hubeny:2007xt}, which generates the evolution of initial data on the Cauchy surface, interpreted as a kink transformation \cite{Kaplan:2022orm}. The action of HRT area on the initial data also preserves the constraint equations, which yields some equations for the RT profile $\rho(x)$. This is also a reformulation of the location of extremal surface \cite{Bousso:2020yxi}.

\subsection{The shockwave method}
In this subsection, we derive the equations for the shockwave profile $h(x)$ from Hamiltonian formalism. As we mentioned in the introduction and section \ref{sec:butterfly v}, we first consider the eternal black hole (\ref{eternal BH}) in AdS$_{d+1}$. To manifest the foliation of the spacetime, we set null direction by
\be\label{uv} u=T-\rho,\quad v=T+\rho, \ee
then take a pullback of (\ref{eternal BH}) by (\ref{uv}) as
\be\label{eternal BH2} ds^2=-a(T^2-\rho^2)(dT^2-d\rho^2)+r(T^2-\rho^2)^2\delta_{ij}dx^idx^j. \ee
For convenience, we focus on a Cauchy surface $\Sigma_0$ with $T=0$. Applying the foliation (\ref{foliation}) and (\ref{Kab}) on (\ref{eternal BH2}) yields the induced metric on the Cauchy surface $\Sigma_0$
\be\label{induced metric1} \sigma_{AB}dY^AdY^B=a(-\rho^2)d\rho^2+r(-\rho^2)^2\delta_{ij}dx^idx^j \ee
and the extrinsic curvature of Cauchy surface $\Sigma_0$
\be\label{KAB1} K_{AB}=0, \ee
by time reversal symmetry of (\ref{eternal BH2}).

We then consider a single shockwave near $u=0$ such that the backreacted metric $\widetilde{g}_{\mu\nu}=g_{\mu\nu}+\D g_{\mu\nu}$ is (\ref{singlesw}), where the action of single shockwave is
\be\label{evo singlesw} \D g_{\mu\nu}=h(x)\d (u)\d_\mu^u \d_\nu^u. \ee
We also take a pullback of (\ref{singlesw}) by (\ref{uv}) as
\be\label{singlesw2} ds^2=-a(T^2-\rho^2)(dT^2-d\rho^2)+r(T^2-\rho^2)^2\delta_{ij}dx^idx^j+h(x)\delta(T-\rho)(dT^2-2dTd\rho+d\rho^2). \ee
Applying foliation (\ref{foliation}) and (\ref{Kab}) on (\ref{singlesw2}) yields the evolved induced metric on the Cauchy surface $\Sigma_0$
\be\label{induced metric2} \widetilde{\sigma}_{AB}dY^AdY^B=a(-\rho^2)d\rho^2+r(-\rho^2)^2\delta_{ij}dx^idx^j+h(x)\delta(\rho)d\rho^2, \ee
and the evolved extrinsic curvature $\widetilde{K}_{AB}$ of Cauchy surface $\Sigma_0$
\be\label{KAB2} \widetilde{K}_{AB}=\frac{1}{2\sqrt{a(0)}}\Big[h(x)\delta^\prime(\rho)\delta^\rho_A\delta^\rho_B+\partial_i h(x)\delta(\rho)(\delta^i_A\delta^\rho_B+\delta^\rho_A\delta^i_B)\Big]. \ee
We present the detailed calculation of $\widetilde{K}_{AB}$ in Appendix. \ref{sec:Appen}. Combining (\ref{induced metric1}), (\ref{KAB1}), (\ref{induced metric2}) and (\ref{KAB2}), we present the evolution of initial data on the Cauchy surface $\Sigma_0$ under the single shockwave (\ref{evo singlesw}) as
\begin{align}\label{evo id1} 
	\Delta\sigma_{AB}=&\widetilde{\sigma}_{AB}-\sigma_{AB}=h(x)\delta(\rho)\delta^\rho_A\delta^\rho_B\notag\\
	\Delta K_{AB}=&\widetilde{K}_{AB}-K_{AB}=\frac{1}{2\sqrt{a(0)}}\Big[h(x)\delta^\prime(\rho)\delta^\rho_A\delta^\rho_B+\partial_i h(x)\delta(\rho)(\delta^i_A\delta^\rho_B+\delta^\rho_A\delta^i_B)\Big].
\end{align}

Given the evolution of initial data (\ref{evo id1}), we can apply it to the constraint equations
\begin{align}
	\mathcal{H}=&\frac{1}{2}\sqrt{\sigma}\Big(R^{(d)}+K^2-K^{AB}K_{AB}+d(d-1)\Big)\notag\\
	\mathcal{H}_A=&\sqrt{\sigma}D^B(K_{AB}-K\sigma_{AB}),
\end{align}
where $R^{(d)}$ is the Ricci scalar of $\sigma_{AB}$, $K$ is the trace of the extrinsic curvature \footnote{Strictly speaking, $K$ is the extrinsic curvature, while $K_{AB}$ is the second fundamental form. But we also name $K_{AB}$ the extrinsic curvature to emphasize its geometric meaning in this paper.}, $K=\sigma^{AB}K_{AB}$, $D$ denotes the covariant derivative associated with $\sigma_{AB}$. Considering that the evolution of initial data (\ref{evo id1}) preserves the constraint equations
\begin{align}
	\mathcal{H}=&0\notag\\
	\mathcal{H}_A=&0,
\end{align}
which yields an equation for the shockwave profile $h(x)$
\be\label{h Hamiltonian} \Big(\frac{1}{2r^2}\delta^{ij}\partial_i\partial_jh(x)-\frac{1}{4}d(d-1)h(x)\Big)\delta(\rho)=0,
\ee
which manifests in (\ref{DH}) and (\ref{DHA}).

\begin{figure}[h]
\centering
\includegraphics[width=6cm]{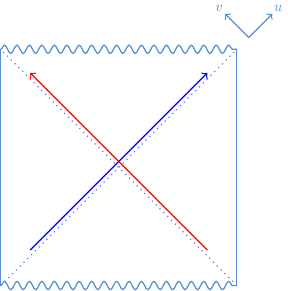}
\caption{An interpretation of double shockwave: two infalling particles into black hole.}\label{Fig:double shockwave}
\end{figure}

Note that the single shockwave brings a non-vanishing change for the induced metric (\ref{evo id1}), i.e., $\Delta\sigma_{AB}\neq0$, which is not consistent with the picture of entanglement wedge reconstruction, where the action of the HRT area can not change the induced metric. To remedy this discrepancy and make the intrinsic structure of Cauchy surface consistent with entanglement wedge reconstruction, we can introduce a double shockwave to cancel out this non-vanishing change for the induced metric such that $\Delta\sigma_{AB}=0$. More precisely, we just add an extra single shockwave near $v=0$ in (\ref{singlesw}) such that the backreaction of double shockwave is
\be\label{doublesw}
ds^2=-a(uv)dudv + r(uv)^2\delta_{ij}dx^idx^j+h(x)\delta(u)du^2 - h(x)\delta(v)dv^2,
\end{equation}
as shown in Fig.\ref{Fig:double shockwave}.
Notice that there are just a spatial reflection $\rho\to -\rho$ and a reverse profile $h(x)\to -h(x)$ for the two single shockwave on $\Sigma_0$. Therefore, from (\ref{evo id1}), we can easily get the evolution of initial data on $\Sigma_0$ under the double shockwave (\ref{doublesw})
\begin{align}\label{evo id2}
\Delta\sigma_{AB}&=0\notag\\
\Delta K_{AB}&=\frac{1}{a(0)^{\frac{1}{2}}}(h(x)\delta'(\rho)\delta^{\rho}_A\delta^{\rho}_B+\partial_ih(x)\delta(\rho)(\delta^{\rho}_A\delta^{i}_B+\delta^{i}_A\delta^{\rho}_B)),
\end{align}
and the evolution of constraint equations under the double shockwave (\ref{doublesw})
\begin{align}\label{DH dw}
\Delta\mathcal{H}&=0\notag\\
\Delta\mathcal{H}_A&=\frac{2}{a^{\frac{1}{2}}(0)}\Big(\frac{1}{2r^2}\delta^{ij}\partial_i\partial_jh(x)-\frac{1}{4}d(d-1)h(x)\Big)\delta(\rho)\delta^\rho_A,
\end{align}
which also yields the equation (\ref{h Hamiltonian}) for the shockwave profile $h(x)$.

\subsection{The entanglement wedge method}\label{subsec:EW Hamil}
In this subsection, we derive the equations for the RT profile $\rho(x)$ from Hamiltonian formalism. Before proceeding to the entanglement wedge reconstruction method, we review a less used way \cite{Bousso:2020yxi} to determine the location of extremal surface. Instead of extremizing the entanglement entropy, e.g., the HRT area, to find the extremal surface, here we use the gravitational constraint equations. The intuition is that the extremal surface under some evolution of the initial data is still an extremal surface if and only if this evolution preserves the constraint equations.

\begin{figure}[h]
	\centering
	\includegraphics[width=8cm]{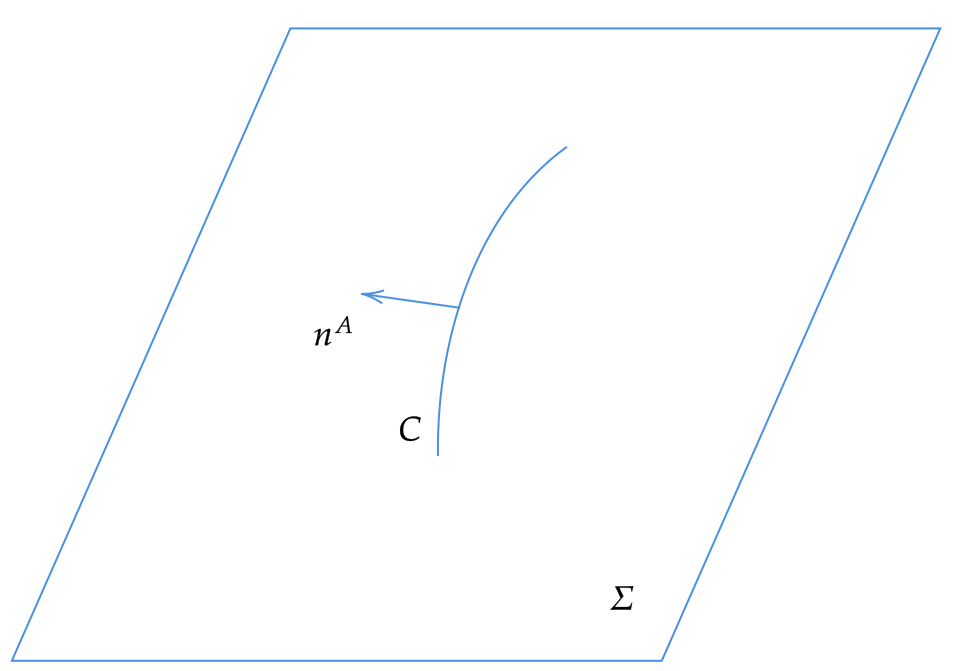}\\
	\caption{An illustration of an extremal surface and Cauchy surface.}\label{setup of extremal surface}
\end{figure}

For example, we consider a codimension two extremal surface $C$ and a Cauchy surface $\Sigma$ that includes $C$, as shown in Fig. \ref{setup of extremal surface}. We construct an evolution of initial data
\begin{align}\label{evo id3}
	\Delta\sigma_{AB}&=0 \notag\\
	\Delta K_{AB}&=\lambda\cdot 2\pi\d(\rho)n_an_b,
\end{align}
where $\lambda$ is an infinitesimal parameter, $\rho$ is a function on the Cauchy surface denoting the distance to the extremal surface $C$, $n_A$ denotes a covector to extremal surface on the Cauchy surface. We apply (\ref{evo id3}) on the constraint equations
\begin{align}
C_1=&\frac{1}{16\pi G}(\widetilde{R}+K^2-K_{AB}K^{AB}+d(d-1))\notag\\
C_{2,A}=&\frac{1}{8\pi G}(D^BK_{AB}-D_AK),
\end{align}
where $\widetilde{R}$ denotes the Ricci scalar associated with the induced metric $\sigma_{AB}$. It yields the evolution of constraint equations
	\begin{align}
\D C_1=&\frac{\lambda}{4G}(\sigma^{mn}-n^mn^n)K_{mn}\d(\rho)\notag\\
\D C_{2,a}=&\frac{\lambda}{4G}(n^mD_mn_a+\sigma^{mn}n_aD_mn_n)\d(\rho),
\end{align}
which yields the extremality of $C$ in spacetime
\begin{align}
\D C_1=&\frac{\lambda}{4G}h^{\mu\nu}\nabla_\mu\tau_\nu\d(\rho)\notag\\
n^a\D C_{2,a}=&\frac{\lambda}{4G}h^{\mu\nu}\nabla_\mu n_\nu\d(\rho).
\end{align}
Here $h_{\mu\nu}$ denote the induced metric on extremal surface $C$. We have revisited holding the constraint equations is equivalent to holding the extremality of $C$ \cite{Bousso:2020yxi}, \be \D C_1=\D C_{2,a}=0 \Leftrightarrow h^{\mu\nu}\nabla_\mu\tau_\nu=h^{\mu\nu}\nabla_\mu n_\nu=0. \ee 

\begin{figure}[h]
  \centering
  \includegraphics[width=6cm]{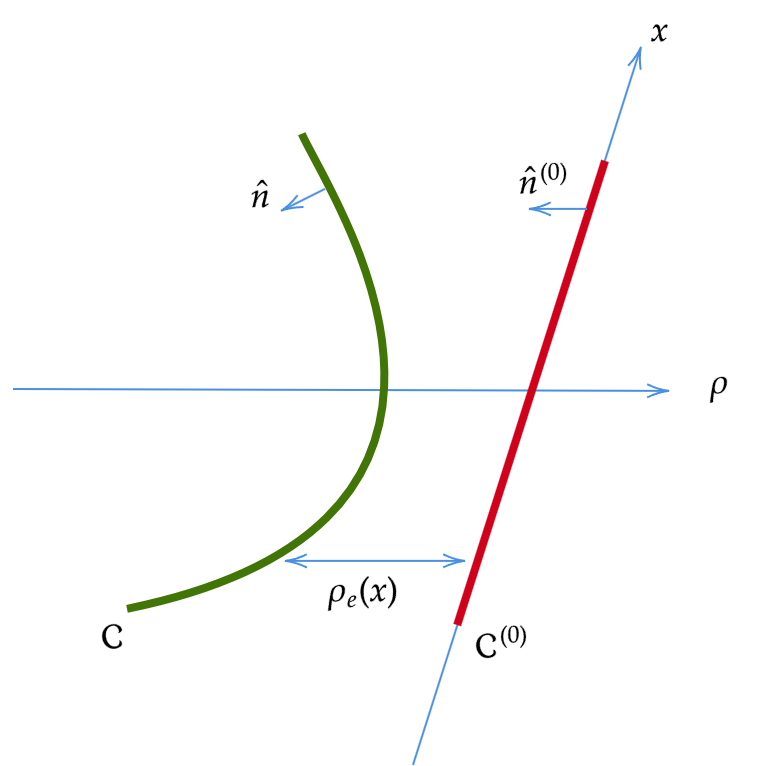}
  \caption{An illustration of our setup. The green line denotes the extremal surface. The red line denotes a reference surface (also the horizon).}\label{EW construction}
\end{figure}

Now we start to derive the equations for the RT profile $\rho(x)$. We first locate the extremal surface $C$ on $\Sigma_0$ by considering the following evolution of initial data
\begin{align}
\Delta\sigma_{AB}&=0\notag\\
\Delta K_{AB}&=k\delta(s(Y))n_An_B,
\end{align}
where the location of extremal surface is denoted by $s(Y)=0$ on $\Sigma_0$. For convenience, we can construct a reference surface $C_0$ (also denoting the horizon) at $\rho=0$, and this is equivalent to adding an extra term in the initial value which has no contribution in the constraint equations
\be\label{dK} \Delta K_{AB}=k\Big(\delta(s(Y))n_An_B-\delta(s^{(0)}(Y))n^{(0)}_An^{(0)}_B\Big). \ee
In the near horizon region, we set the gap between $C$ and $C_0$ to be the RT profile $\rho_e(x)$ such that $\rho_e(x)$ is small enough, as shown in Fig. \ref{EW construction}.

In this setup, we can give a geometric interpretation for quantities appearing in (\ref{dK}). More precisely, we set 
\begin{align}\label{sY}
s^{(0)}(Y)=&\sqrt{a}(\rho-0)\notag\\
s(Y)=&\sqrt{a}(\rho-\rho_e(x))
\end{align}
and
\begin{align}\label{nA}
n^{(0)}_\rho=&\sqrt{a},\quad n^{(0)}_i=0\notag\\
n_\rho=&\sqrt{a},\quad n_{i}=-\sqrt{a}\partial_i\rho_e(x)
\end{align}
such that 
\begin{align}
    n^{(0)}_A|_{C_0}=&\partial_A s^{(0)}(Y)|_{C_0}\notag\\
    n_A|_{C}=&\partial_A s(Y)|_{C}.
\end{align}
Here (\ref{sY}) further yields
\begin{align}\label{del sY}
\delta(s^{(0)}(Y))=&\frac{1}{\sqrt{a}}\delta(\rho)\notag\\
\delta(s(Y))=&\frac{1}{\sqrt{a}}\delta(\rho-\rho_e(x))=\frac{1}{\sqrt{a}}\delta(\rho)-\frac{1}{\sqrt{a}}\rho_e(x)\delta^\prime(\rho).
\end{align}
Applying (\ref{nA}) and (\ref{del sY}) on (\ref{dK}) yields
\be \D K_{AB}=-ka(0)^{\frac{1}{2}}\Big(\rho_e(x)\delta'(\rho)\delta^{\rho}_A\delta^{\rho}_B+\partial_i\rho_e(x)\delta(\rho)(\delta^{\rho}_A\delta^{i}_B+\delta^{i}_A\delta^{\rho}_B)\Big). \ee
So far we have obtained the evolution of initial data in the entanglement wedge reconstruction method
\begin{align}\label{evo id4}
	\D \sigma_{AB}=&0\notag\\
	\D K_{AB}=&-ka(0)^{\frac{1}{2}}\Big(\rho_e(x)\delta'(\rho)\delta^{\rho}_A\delta^{\rho}_B+\partial_i\rho_e(x)\delta(\rho)(\delta^{\rho}_A\delta^{i}_B+\delta^{i}_A\delta^{\rho}_B)\Big).
\end{align}
Applying (\ref{evo id4}) on constraint equations yields an equation for the RT profile $\rho_e(x)$
\be\label{rho Hamiltonian} \Big(\frac{1}{2r^2}\delta^{ij}\partial_i\partial_j\rho_e(x)-\frac{1}{4}d(d-1)\rho_e(x)\Big)\delta(\rho)=0.
\ee
\subsection{Equivalence}
Based on the evolution of initial data with two distinct methods, we demonstrate the equivalence of two butterfly velocity computations in this subsection. We now collect our previous initial data structure here
\begin{align}
\Delta\sigma_{AB}=&0\notag\\
\Delta K_{AB}=&\frac{1}{a(0)^{\frac{1}{2}}}\Big(h(x)\delta'(\rho)\delta^{\rho}_A\delta^{\rho}_B+\partial_ih(x)\delta(\rho)(\delta^{\rho}_A\delta^{i}_B+\delta^{i}_A\delta^{\rho}_B)\Big)
\end{align}
for the double shock wave method, and
\begin{align}
\Delta\sigma_{AB}=&0\notag\\
\Delta K_{AB}=&-ka(0)^{\frac{1}{2}}\Big(\rho_e(x)\delta'(\rho)\delta^{\rho}_A\delta^{\rho}_B+\partial_i\rho_e(x)\delta(\rho)(\delta^{\rho}_A\delta^{i}_B+\delta^{i}_A\delta^{\rho}_B)\Big)
\end{align}
for the entanglement wedge method. It shows the same algebraic structure for both $h(x)$ and $\rho_e(x)$.

Furthermore, inserting above initial data into the constraint equations, we can obtain same equations for $h(x)$ and $\rho_e(x)$,
\begin{align}
\Big(\frac{1}{2r^2}\delta^{ij}\partial_i\partial_jh(x)-\frac{1}{4}d(d-1)h(x)\Big)\delta(\rho)=&0\notag\\ \Big(\frac{1}{2r^2}\delta^{ij}\partial_i\partial_j\rho_e(x)-\frac{1}{4}d(d-1)\rho_e(x)\Big)\delta(\rho)=&0.
\end{align}
Thus we demonstrate the equivalence of two butterfly velocity computations without explicitly solving these equations for $h(x)$ and $\rho_e(x)$.

\section{Conclusion and Discussion}\label{sec:discussion}

We have revisited the equivalence of two holographic methods in computing butterfly velocity from the Hamiltonian formalism of gravity, especially from the same algebraic structure for the initial data appearing in both methods.

We now comment on some potential future directions:

Inspired by the algebraic structure of the extrinsic curvature, 
\be K_{AB} = \lambda(x)\delta'(\rho)\delta_A^\rho\delta_B^\rho + \partial_i\lambda(x)\delta(\rho)(\delta_A^\rho\delta_B^i + \delta_A^i\delta_B^\rho), \ee
we may implement a time direction split scheme parameterized by $\lambda(x)$. 
\begin{figure}[h]
  \centering
  \includegraphics[width=6cm]{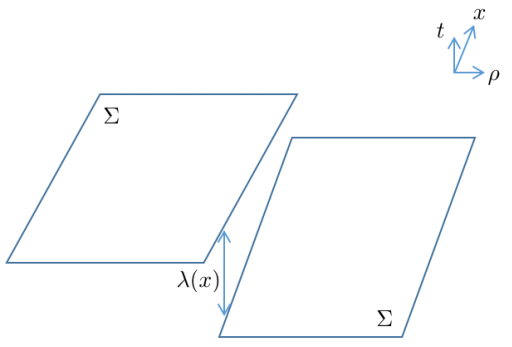}
  \caption{An illustration of time direction split scheme.}\label{timesplit}
\end{figure}
It manifests in both methods. In double shock wave method, a spacelike geodesic running from the right to the left crosses both shocks and receives a discontinuous shift at each interface. The combined shifts produce a net displacement in the time direction, giving a geometric picture of how the shocks split the Cauchy surface in time. In entanglement wedge reconstruction method, the term $\delta(s)n_{A}n_{B}$ corresponds to a Minkowski conical defect. It characterizes a time direction shift between the two sides of extremal surface.

We revisit this equivalence only in pure AdS$_{d+1}$ gravity in this work. Nevertheless, we find that, from the Hamiltonian formalism perspective, the backreaction of infalling particle that is not manifest in entanglement-wedge reconstruction is encoded in action of diffeomorphism-invariant observable on the Cauchy surface. If we attempt to extend this approach to other gravitational systems—for example, higher-derivative gravity theories—we will need appropriate diffeomorphism-invariant observables, about which our current understanding remains very limited. Furthermore, considering a gravitationally dressed particle creation operator in gravity may better demonstrate the equivalence of two holographic methods.

In addition, Jafferis-Lewkowycz-Maldacena-Suh (JLMS) formula \cite{Jafferis:2015del} establishes that the gravitational area operator is dual to the modular Hamiltonian in CFT. By evolving the system forward and backward with different modular Hamiltonian $e^{-iH_B s} e^{iH_A s} \rho e^{-iH_A s} e^{iH_B s}$, one can read out a lot of universal properties of the field theory, for example
the quantum null energy condition \cite{Balakrishnan:2017bjg}. The extrinsic curvature  \( K_{AB} = k \left( \delta(s(Y)) n_A n_B - \delta(s^{(0)}(Y)) n_A^{(0)} n_B^{(0)} \right) \) we construct is precisely a forward plus a backward evolution by two area observable. It will be interesting to figure out their relation.

\section*{Acknowledgments}

We thank for the discussion with Banashree Baishya, Hong Liu, Farzad Omidi, Baishali Roy, Elisa Tabor, Jun-Kun Zhao. X.s.W thanks the 19th Asian Winter School on Strings, Particles and Cosmology. X.s.W thanks the Yukawa Institute for Theoretical Physics at Kyoto University, where this work was initiated partially done during the workshop on "Black Hole, Quantum Chaos and Quantum Information". X.s.W thanks the China-India-UK School in Mathematical Physics, held by ICMS, Edinburgh.

X.s.W. and J.q.W. are supported by the National Natural Science Foundation of China (NSFC) Project No.12447101.

\appendix

\section{More details on Hamiltonian formalism}\label{sec:Appen}
In this appendix, we provide more details on Hamiltonian formalism used in section \ref{sec:Hamiltonian}.

We first recall that we foliate the spacetime with a set of Cauchy surfaces denoted by a scalar field $T$ such that we can present the metric as
\be ds^2=-N^2dT^2+\sigma_{AB}(dX^A+\beta^Adt)(dX^B+\beta^Bdt), \ee
where $N$ is lapse function, $\hat{\beta}$ is shift vector and $\sigma_{AB}$ is the induced metric on the Cauchy surface. Given a metric, e.g., (\ref{eternal BH2}) or (\ref{singlesw2}), one can directly read off $N,\hat{\beta}$ and $\sigma_{AB}$. We introduce the covector $\tau_\mu$ to the Cauchy surface, defined by 
\be \tau_\mu=-N\d^t_\mu, \ee
e.g., the components of $\hat{\tau}$ for (\ref{eternal BH2})
\be\label{tau1} \tau_T=-\sqrt{a(T^2-\rho^2)},\quad \tau_\rho=\tau_i=0. \ee

We next compute the first quantity that is not straightforward to evaluate, i.e., the extrinsic curvature $\widetilde{K}_{AB}$ of $\Sigma_0$ under the backreacted metric (\ref{singlesw2}). Here we can focus on the contribution of $\D g_{\mu\nu}$ (\ref{evo singlesw}) to $\widetilde{K}_{AB}$ since $K_{AB}=0$ by time reversal symmetry of (\ref{eternal BH2}), i.e.,
\be \D K_{AB}=\widetilde{K}_{AB}-K_{AB}=\widetilde{K}_{AB}. \ee
Based on the definition of extrinsic curvature (\ref{Kab}), we present $\Delta K_{AB}$ as
\be\label{evo KAB} \Delta K_{AB}=\Delta\sigma^\mu_A\nabla_\mu\tau_B+\sigma^\mu_A\nabla_\mu\Delta\tau_B-\sigma^\mu_A\Delta\Gamma^\nu_{\mu B}\tau_\nu, \ee
where evolved quantities can be presented by the evolved metric $\D g_{\mu\nu}$ 
\begin{align}\label{evo folia}
	\Delta\tau_\mu=&-\frac{1}{2}\tau^\nu\tau^\rho\Delta g_{\nu\rho}\tau_\mu\notag\\
	\Delta\Gamma^\rho_{\mu\nu}=&\frac{1}{2}g^{\rho\lambda}(\nabla_\mu\Delta g_{\nu\lambda}+\nabla_\nu\Delta g_{\mu\lambda}-\nabla_\lambda\Delta g_{\mu\nu})\notag\\
	\Delta\sigma^\mu_A=&\Delta\tau_A\tau^\mu+\tau_A\tau_\nu\Delta g^{\mu\nu}+\tau_Ag^{\mu\nu}\Delta\tau_\nu.
\end{align}
Applying (\ref{tau1}) and (\ref{evo singlesw}) on (\ref{evo KAB}) yields\footnote{An alternative way to quickly check these results is by taking another equivalent definition of extrinsic curvature
	\be \Delta K_{AB}=-\frac{1}{2}(\mathcal{L}_{\Delta\tau}g_{AB}+\mathcal{L}_\tau\Delta g_{AB}). \ee}
\begin{align}\label{evo KAB1}
	\Delta K_{AB}=&-\frac{1}{2}g^{TT}(\partial_A\Delta g_{BT}+\partial_B\Delta g_{A T}-\partial_T\Delta g_{AB})\tau_T\notag\\
	=&\frac{1}{2\sqrt{a(0)}}\Big[\delta^\rho_A\delta^\rho_B h(x)\delta^\prime(\rho)+(\delta^i_A\delta^\rho_B+\delta^\rho_A\delta^i_B)\partial_i h(x)\delta(\rho)\Big].
\end{align}

We then consider the constraint equations, i.e., the scalar constraint $\mathcal{H}$ and vector constraint $\mathcal{H}_A$
\begin{align}
	\mathcal{H}=&\frac{1}{2}\sqrt{\sigma}\Big(R^{(d)}+K^2-K^{AB}K_{AB}+d(d-1)\Big)\notag\\
	\mathcal{H}_A=&\sqrt{\sigma}D^B(K_{AB}-K\sigma_{AB}),
\end{align}
where one can easily generalize our previous derivation in AdS$_3$ in \cite{Wang:2023jfr} to AdS$_{d+1}$. We can see that the constraint equations are presented by the set of initial data. Applying (\ref{induced metric1}) and (\ref{KAB1}) on scalar constraint equation yields a constraint on the Cauchy surface $\Sigma_0$ under (\ref{eternal BH2})
\be\label{constraintH}	\mathcal{H}\sim R^{(d)}+d(d-1)=\frac{4d}{a^2r^2}\Big[\rho^2\Big(a^\prime rr^\prime-(d-1)a(r^\prime)^2-2arr^{\prime\prime}\Big)+arr^\prime\Big]+d(d-1)=0. \ee
Given the evolution of initial data (\ref{evo id1}) under the single shockwave (\ref{evo singlesw}), we can reformulate the evolution of constraint equations into
\begin{align}\label{DH}
	\frac{1}{\sqrt{\sigma}}\Delta\mathcal{H}=&\frac{1}{\sqrt{\sigma}}\Delta\Big(\frac{1}{2}\sqrt{\sigma}[R^{(d)}+K^2-K^{AB}K_{AB}+d(d-1)]\Big)\notag\\
	=&\frac{1}{2\sqrt{\sigma}}\Big(\sqrt{\sigma}\Delta R^{(d)}+\frac{1}{2}\sqrt{\sigma}\sigma^{AB}\Delta\sigma_{AB}d(d-1)\Big)\notag\\
	=&-\frac{1}{a}\Big(\frac{1}{2r^2}\delta^{ij}\partial_i\partial_jh(x)-\frac{1}{4}d(d-1)h(x)\Big)\delta(\rho)
\end{align}
and
\begin{align}\label{DHA}
	\frac{1}{\sqrt{\sigma}}\Delta\mathcal{H}_A=&\frac{1}{\sqrt{\sigma}}\Delta(\sqrt{\sigma}D^B(K_{AB}-K\sigma_{AB}))\notag\\
	=&\frac{1}{2\sqrt{a}r^2}\delta^{ij}\partial_i\partial_jh(x)\delta(\rho)\delta^\rho_A-\frac{r^\prime d}{a^{\frac{3}{2}}r}\rho h(x)\delta^\prime(\rho)\delta^\rho_A\notag\\
	=&\frac{1}{a^{\frac{1}{2}}}\Big(\frac{1}{2r^2}\delta^{ij}\partial_i\partial_jh(x)-\frac{1}{4}d(d-1)h(x)\Big)\delta(\rho)\delta^\rho_A,
\end{align}
where the derivation is similar to our previous derivation in AdS$_3$ \cite{Wang:2023jfr}, and we have used the non-vanishing components for connection on Cauchy surface $\Sigma_0$
\begin{align}
	\widetilde{\Gamma}^\rho_{\rho\rho}=&-\frac{\rho}{a(-\rho^2)}a^\prime(-\rho^2)\notag\\
	\widetilde{\Gamma}^\rho_{ij}=&\frac{2r(-\rho^2)r^\prime(-\rho^2)}{a(-\rho^2)}\rho\delta_{ij}\notag\\
	\widetilde{\Gamma}^i_{\rho j}=&-\frac{2r^\prime(-\rho^2)}{r(-\rho^2)}\rho\delta^i_j,
\end{align}
and its evolution under the single shockwave (\ref{evo singlesw})
\begin{align}
	\Delta\widetilde{\Gamma}^\rho_{\rho\rho}=&\frac{1}{2a}h(x)\delta^\prime(\rho)\notag\\
	\Delta\widetilde{\Gamma}^i_{\rho\rho}=&-\frac{1}{2r^2}\partial_ih(x)\delta(\rho)\notag\\
	\Delta\widetilde{\Gamma}^\rho_{\rho i}=&\frac{1}{2a}\partial_ih(x)\delta(\rho),
\end{align}
and in deriving the last equation of (\ref{DHA}), we also have used scalar constraint (\ref{constraintH}).

\end{document}